# EQUATION OF STATE OF FERMIONS IN NEUTRON STARS

Samina Masood and Serkan Caliskan
*Department of Physical and Applied Sciences*
*University of Houston-Clear Lake*
*Houston, TX 77058*

## Abstract

Employing both the number density and energy density, the equation of state of fermions for each microscopic region of a neutron star is derived in the presence of various statistical effects. It is computed as functions of temperature ($T$), chemical potential (\mu) and magnetic field ($B$) based on the statistical properties and corresponding phases of neutron stars. It is revealed that the measurable properties describing the dynamic behavior of a neutron star are determined by the unusual statistical conditions. It is demonstrated that, in a strong magnetic field ($B \gg$ \mu), the pressure ($P$) shows a logarithmic dependence on $B$ ($P$ \propto ln (\mu/B) ). It is found that the regions with a higher (lower) magnetic field (density) have a shorter (longer) life time. Moreover, the relationship between the magnetic field and chemical potential confirms that the large magnetic field and high density are short lived as compared to low magnetic field and low density at the same temperature.

## 1. Introduction

Neutron stars (NSs) are extremely dense, rapidly rotating stars with very strong magnetic field. These faint objects have peculiar properties [1-2] and are usually created as a result of supernova explosion and then converted into a white dwarf or merge with another neutron star to create a black hole. A combination of extremely high densities, very hot temperatures with the largest known magnetic field of rapidly rotating neutron stars make them worth-investigating exotic systems incorporating the statistical background effects. In addition to that, these hardly detectable objects emit a small range of frequencies, especially in the visible spectrum. Due to the emission of selective radiation, understanding of the composition and structure becomes even more challenging. On the other hand, dynamics of these relativistic systems cannot be fully understood without considering quantum statistical effects of such rarely found systems. All particles and radiations in neutron stars interact among themselves at extremely high densities and strong magnetic fields which lead to the coexistence of matter in various phases simultaneously. These phases may be limited to small bubbles due to the presence of short-range forces and extreme statistical conditions.

Due to the high baryonic densities these stars are expected to be a home of rarely found matter phases, enclosed in microgranules, such as superfluid and quark-gluon plasma. However, its detailed unusual structure (formed by microgranules) and composition are not fully known yet. Most of the existing stellar models are based on central forces like gravity and electromagnetic force, whereas the major forces in neutron stars are nuclear forces including electroweak and strong interactions. They create structural patterns inside, which may not necessarily be spherical due to lack of central forces. These localized structural patterns may exist in different phases with various statistical properties and composition. Due to individual angular momentum, they exhibit extremely strong local magnetic field that is not found anywhere else. However, the magnetic field creation in small bubbles may vary due to difference of composition and rotational properties. The quantum magnetic collapse [3] can place locally creating huge magnetic field. The existence of weak magnetic field in the vicinity of strong magnetic field and highly squeezed matter and the magnetic field gradient in superdense stars not only traps most of the electromagnetic signals [4] but also create favorable conditions for superfluid [5] during the burning of hydrogen into helium. Such objects are then detected by highly energetic gamma rays and by the strong magnetic field.

Employing the microscopic theory of nuclear matter, a NS can be described by means of equation of state (EOS) which can be expressed in terms of pressure, number density and temperature, in general. Derivation of EOS [6, 7] for both the core and crust has been attracted extensive interest. The works on the EOS are still attractive due to their unique structure and unusual parameters which are needed to be incorporated. They might be revealed from observations or relevant experiments [8]. Obtaining the EOS in an appropriate form for, especially, hot, and dense matter plays a crucial role for simulations and visualizations. A very high density in the inner core and thermal effects are challenging in getting the EOS, which require new techniques and relevant approaches.

For a better understanding of structure and composition of NSs we need to know the modifications in the properties of propagating particles in superdense media, which has already been investigated in literature for electromagnetically interacting systems [9]. That study can be extended to include a strong magnetic field in a relatively straightforward way to be applicable to various regions of a NS. In the present paper, we develop an EOS for each microscopic region with unusual conditions in a NS based on their statistical properties and corresponding phases. Each proposed layer of a NS can be considered as a heterogeneous layer that is made up of small granules of various phases exhibiting distinct local statistical conditions. Overall structure of a NS is built based on known physical processes and observational data. EOS deals with overall parameters of microscopic granules. The overall parameters associated with various layers are determined as an integrated effect of individual particle behavior due to the statistical effects. Therefore, we need to determine the EOS for every region to understand its overall dynamic behavior.

In Section 2, we briefly describe the calculational scheme to develop the appropriate EOS for fermions in various regions of NSs. In Section 3, we discuss the motivation to derive an EOS corresponding to possible phases of matter in NSs. The EOS is expressed for the given statistical conditions and its relevance with NSs is discussed in the last section. In the end, a possible model for a NS is proposed.

## 2. Calculational Scheme

Extremely high temperatures and densities of compact stellar objects indicate relativistic motion of individual particles that needs to be described by quantum field theory which is used to study the dynamics of such systems. In this paper, we study the effects of statistical conditions on the propagation of leptons using Quantum Electrodynamics (QED) and electroweak theory because they are the highly relativistic particles in such media due to their light mass. We use the renormalization scheme of QED in hot and dense media with or without magnetic field. Strong interaction is a characteristic of hydronic matter and is not realized by leptons and can be easily ignored. Study of renormalization is only possible in the real-time formalism [10,11] where order-by-order cancellation of singularities can be proven, explicitly. Each propagating particle has nonzero probability to interact with the surrounding particles, which are abundantly produced in extreme statistical conditions and modify the propagator in a Feynman diagram because of its interaction with the particles in the medium. The real-time formalism is used here as it provides an opportunity to compute background contribution to physically measurable parameters. We consider an electrodynamically interacting medium of electrons, positrons and the electromagnetic radiations. All of these particles are in thermal equilibrium. We use the statistical conditions [6] already determined based on some observations and proposed by some existing models. We can use the renormalized parameters as effective parameters of the theory. These renormalized QED parameters give some additional information about the structure and composition of stars. Fermi-Dirac and Bose-Einstein distribution functions are integrated over the phase space to get the number density of the corresponding particles in a system incorporating the possibility of interaction with the medium. Interactions with electrons, positrons and photons in a hot and dense medium are included in the theory by adding statistical distribution functions in the particle propagators [12]. Electromagnetic interactions of particles are modified in this background because many-body aspects are possessed by the statistical media and replace the individual particle behavior in vacuum by the particle distribution functions in a statistical ensemble. Feynman rules of statistical field theory remain the same as those in vacuum except that the particle propagators have to be appropriately modified. Renormalization techniques of vacuum theory are extended to include finite temperature and density effects using the standard statistical distribution functions of fermions and bosons in the corresponding particle propagators. We work in a space where the Green's functions depend on real Minkowski momenta. Therefore, the dynamical processes such as particles propagating in the heat-bath may be more conveniently dealt within the real-time formalism [11-15]. In this calculational scheme, the fermion propagator in a QED medium becomes a function of Fermi-Dirac distribution function due to its possible interaction with the medium radiatively [4]. In this case, we use the relativistic representation of 4-momentum in terms of energy and momentum of fermions of mass $m$ in 3-dimensional space and a strong magnetic field responsible of Landau distribution in $l$ Landau energy levels;

$$E^2 = p^2 + m^2 + eB(2l + 1) = p^2 + M^2 \tag{1}$$

where, for the electron mass $m$, the corresponding relativistic energy can be expressed by the magnetic field $B$ of $10^9$ T. On the other hand, the magnetic field of a NS can go up to $10^{12}$ T or even higher for magnetars [3]. However, incorporating the effect of magnetic field is not necessary unless the magnetic field in a relativistic system does not approach close to $10^9$ T. The high magnetic field is definitely applicable to high $B$ limits of NSs, especially Magnetars. It can still be below $10^{12}$ T where quantum magnetic collapse can take place [3]. Employing the $M$, the magnetic field $B$ can be taken into account through the interaction of electron with the magnetic field (the effective mass of the electron rises significantly due to its coupling with a high magnetic field). From Eq. (1), $M = m\left[1 + \frac{eB}{m^2}(2l + 1)\right]^{1/2}$ is related to the $B$ of $10^9$ T associated with the $m$ as follows;

$$\text{for } B \ll m \qquad M = m \tag{2a}$$

$$\text{for } B \sim m \qquad M = m\left[1 + \frac{eB}{m^2}(2l + 1)\right]^{1/2} \tag{2b}$$

$$\text{for } B \gg m \qquad M = m\left[\frac{eB}{m^2}(2l + 1)\right]^{1/2} \tag{2c}$$

In the extreme case, $\frac{eB}{m^2}(2l + 1) \gg 1$, Eq. (2c) can also be applied for high $l$ and low $B$ values.

The Fermi-Dirac distribution function for the electrons is given by;

$$n_F(E, \mu) = \frac{1}{e^{\beta(E-\mu)} + 1} \tag{3}$$

where $\mu$ is the chemical potential assigned to fermions and $E$ depends on magnitude of 3-momentum $p$. For antiparticles $\mu$ is replaced by $-\mu$. However the concentration of free real positrons or antiparticles is never high enough to contribute to the chemical potential, so the effect of high magnetic field and the two degrees of rotation of electrons take care of the magnetic field generation from both incoming and outgoing electrons. In Eq. (3), a large value of magnetic field should also be compared to $\mu$ for more accurate results.

## 3. The Equation of State (EOS)

The calculation of both number density (*N*) and energy density is the most important step towards getting the EOS for the corresponding regions of a NS. Average number density of electrons in NSs can be calculated from the estimated mass density of electrons divided by the electron mass. Number density of electrons is calculated at finite temperature and density from the distribution function of electron in the corresponding systems. Number density is defined as [13]

$$N = 2\int \frac{d^3 p_0}{(2\pi)^3} n_F(p_0, \mu) \tag{4}$$

where $p_0 = E$ corresponds to the energy of the real particles and is given by Eq. (1) for a relativistic system with $p$, the magnitude of the 3-momentum of the electrons. The equation above can also be used to calculate the number of relativistic electrons in the early universe in the presence of different type of particles. Over the entire momentum space, the EOS of various regions of NSs is computed incorporating the interaction of particles with other particles in a medium. As the electron mass is found to be changed in hot and dense media [14-15], direct integration of distribution functions has to be expanded in two ranges of temperature and density. For the distribution within the granule, the number density of particles can be expressed by the integrals over the energy *E* [13];

$$N = \frac{1}{2\pi^2}\left[I_1 - \frac{M^2}{2} I_2 - \frac{M^4}{8} I_3\right] \tag{5}$$

Here, *M* can be evaluated for various regions of high temperature, high density or large magnetic field. In Eq. (5), $I_1$, $I_2$ and $I_3$ represent the integrals: $I_1 = \int_M^\infty E^2\, n_F(E,\mu) dE$, $I_2 = \int_M^\infty n_F(E,\mu) dE$ and $I_3 = \int_M^\infty \frac{n_F(E,\mu)}{E^2} dE$. On the other hand, the average energy density in a medium is defined by;

$$\varepsilon = N\langle E\rangle = \int \frac{p_0 d^3 p}{(2\pi)^3} n_F(p_0, \mu) \tag{6}$$

Employing the number density, it can be expressed in the following form [13];

$$\varepsilon = \frac{1}{2\pi^2}\left[I_4 - \frac{M^2}{2} I_5 - \frac{M^4}{4} I_6\right] \tag{7}$$

where $I_4 = \int_M^\infty E^3 n_F(E,\mu)dE$, $I_5 = \int_M^\infty E n_F(E,\mu)dE$ and $I_6 = \int_M^\infty \frac{n_F(E,\mu)}{E} dE$. The $I_1$ - $I_6$ integrals are given in the general form in Appendix in terms of specific functions that have been evaluated for various regions of statistical conditions relevant to early universe or stellar interiors. Since all the particle species are in thermal equilibrium, it is convenient to expand the $n_F$ in powers of $n\beta$, where $\beta = 1/T$ with $T$ being the temperature. Note that $M$ correspond to the potential energy contribution to the propagating particle. For extremely dense and highly magnetized environment of NSs, $M$ refers to the effect of $B$, and both $T$ and $M$ are expressed in the same units of electron mass or its corresponding energy. The $n_F$ can be converted into the sum by considering the two cases: (i) $\mu >$ both $T$ and $M$, (ii) $\mu <$ both $T$ and $M$;

$$n_F \overset{\mu<T,M}{\rightarrow} \sum_{n=1}^{\infty} (-1)^n e^{n\beta(\mu-p_0)}$$

$$n_F \overset{\mu>T,M}{\rightarrow} \theta(\mu - p_0) + \sum_{n=1}^{\infty} (-1)^n e^{-n\beta(\mu-p_0)}$$

Here $\theta$ is the *step function*, defined as;

$$\theta(p_0) = \begin{cases} 1 & \text{for} \quad p_0 \geq 0 \\ 0 & \text{for} \quad p_0 < 0 \end{cases}$$

Note that $\theta(\mu - p_0)$ cannot exist for antiparticles. The functions (see Appendix) in the integrals acquire different values in different statistical conditions. Upon evaluating the associated integrals, both the $N$ and $\varepsilon$ can be obtained (see Appendix);

$$N = \frac{\mu^3}{2\pi^2}\left[\frac{2}{3}\left(1 - \frac{M^3}{\mu^3}\right) - \frac{M^2}{\mu^2}\left(1 - \frac{M}{\mu}\right) + \frac{M}{\mu}\left(1 - \frac{M^2}{\mu^2}\right) + \frac{M^4}{8\mu^4}\left(1 - \frac{M^3}{\mu^3}\right)\right] \tag{8}$$

$$< P > = \frac{\mu^4}{3\pi^2}\left[\frac{M^3}{4\mu^3}\left(1 - \frac{M}{\mu}\right) + \frac{5M^2}{4\mu^2}\ln\left(\frac{\mu}{M}\right) + \frac{M}{2\mu}\left(1 - \frac{M^3}{\mu^3}\right) - \left(1 - \frac{M^4}{\mu^4}\right)\right] \tag{9}$$

where, using the Fermi gas model, the pressure was obtained through $< P > = \frac{2}{3}\varepsilon$ . Eq. (9) yields a general form of the EOS of state of electrons in NSs and its behavior varies according to the statistical conditions of every envelope of the enclosed material with proper canonical ensemble. Various terms contribute differently in different regions corresponding to the temperature and chemical potential of the domain (i.e., granules). It means that if we know the EOS of electrons, we can reveal statistical behavior of possible local phases of matter in different regions of NSs. Therefore, we may find various granular phases in different layers of NSs. Eq. (9) represents the EOS for a high chemical potential which sometimes, due to rapid rotation, may generate extremely large magnetic fields. It may even rise to dynamo type effect which is possible in magnetars [17]. For small values of $B$, $M \to m$ (electron mass). If $\mu \ll T$, the results are applicable to early universe. However different regions of NSs, especially the core, is a dynamic system with $\mu > T$, on the other hand $T$ is still too high to heat it as a non-relativistic system. If $\mu <$ both $T$ and $M$, applying to Eq. (2), the effect of magnetic field can be exhibited. When $eB(2l+1) \gg m^2$, the comparison of $M^2 \sim eB(2l+1)$ and the lower limits of integrals given by $I_1$ through $I_6$ become modified accordingly: $M \to \sqrt{eB(2l+1)}$ in units of electron mass. Hence, in the presence of $B$, $M$ accounts for the effect of magnetic field in Eq. (9). For an extremely high $\mu$, the Eqs. (8) and (9) above are evaluated for various statistical conditions in the next section.

## 4. Discussion and Conclusion

For a relativistically moving highly dense system with an extremely high chemical potential, as the interactions are strong enough for small regions (granules) only, we expect very small regions to stay within the range of relativistic interactions and validity of Fermi-Dirac distribution. The possibility of layers of granules allows to accept the existence of multiple type of phases in one layer without interfering the possibility of existence of higher to lower overall density layers of matter in NSs. The core has a dominance of nuclear matter whereas the crust can be modelled as an approximate metallic structure to a valid approximation.

The quantum statistical effects in the EOS are deployed by the number density of particles. Precise calculation of $\varepsilon$ incorporating the interaction of propagating particle with a medium allows to modify the average energy in terms of statistical properties. This approach gives information about the structure of NSs and can be harnessed to model them. As the integrals defining the $N$ and $\varepsilon$ are evaluated for $M < p_0 < \infty$, using the uncertainty principle ($\Delta t \sim \frac{\hbar}{\Delta E}$), we can get a lower and upper limit for the lifetime of individual granule: $0 < \Delta t < \frac{\hbar}{M}$. This relation shows the short lifetime of such regions corresponds to the statistical conditions and keep changing rapidly inside NSs indicating an active dynamic behavior of the interior of NSs. It provides an implicit time limit into various phases with particular statistical conditions, implying the *life time* of granules. It shows that different layers of NSs can be composed of small micro sized granules moving around quickly and are related to rapid rotation of NSs, due to the variation in lifetime of individual granules. This dynamic behavior of NSs indicate the coexistence of several phases of matter in each layer of the star. The collective behavior of various layers of NSs is deduced from the individual dynamic behavior of a granule. On the other hand, 3-momentum integrals are performed in 3D space for $0 < |p| < \infty$ which corresponds to the entire region of a NS as $0 < r < \infty$. For the sake of generality, the momentum should be retained within the small regions of granules. Once the life time of granules is calculated in terms of statistical parameters, we can clearly see the effect of these variables on the size and life time of granules. It clearly shows that a high (low) magnetic field can survive for a short (long) time interval. It implies that a higher magnetic field is associated with more active regions of star as compared to highly dense slowly active ones.

Aforementioned equations indicate that in the extreme situation, $B$ is ignorable compared to $\mu$ and can totally be neglected. On the other hand, $\frac{\mu}{T}$ rather than individual parameters $(\mu, T)$ is responsible for the metallic behavior for classical or relativistic systems. Then, if $\mu$ is much higher than $T$ ($\mu \gg T$), Eqs. (8) and (9) reproduces the results for a condensed matter system;

$$N = \frac{\mu^3}{6\pi^2}$$

$$\varepsilon = \frac{\mu^4}{8\pi^2}$$

$$< P > = \frac{\mu^4}{12\pi^2}$$

In the presence of magnetic field, when $M \gg \mu$, if we just consider the first Landau level;

$$N = -\frac{B^4}{16\mu\pi^2}$$

$$\varepsilon = -\frac{B^4}{4\pi^2}\left(1 + \frac{1}{4}\ln\left(\frac{\mu}{B}\right)\right)$$

$$< P > = -\frac{B^4}{6\pi^2}\left(1 + \frac{1}{4}\ln\left(\frac{\mu}{B}\right)\right)$$

The Eqs. above indicate that a low $\mu$ and low $N$ produce small outward pressure (i.e., flux) whereas higher $\mu$ values increase the $< P >$ as $\mu^4$. Therefore, the extremely large chemical potential with ignorable magnetic field will cause a chemical collapse of granule and after certain limit the granule formation cannot occur. On the contrary, a large $B$ contributes significantly to both $< P >$ and $N$, and causes a quantum magnetic collapse for large $B$ values [3,17]. The corresponding behaviors are illustrated in Figs.1a and 1b, employing specific numbers in MeVs. Fig. 1a shows the effect of peak value of pressure due to a particular value of chemical potential. Pressure peak is not only located at different values of magnetic field, it also drops quickly for low chemical potential. Whereas, for large chemical potential, it goes rather quickly to peak values and may not fall back for the magnetic field in range. In Fig. 1b, another interesting fact is noted by plotting pressure against chemical potential for various values of magnetic field. It is observed that the pressure goes to zero for various values of magnetic field for the chemical potential close to 500 MeV, which is providing a natural limit on chemical potential value in the star. The granule should collapse inward for that density.

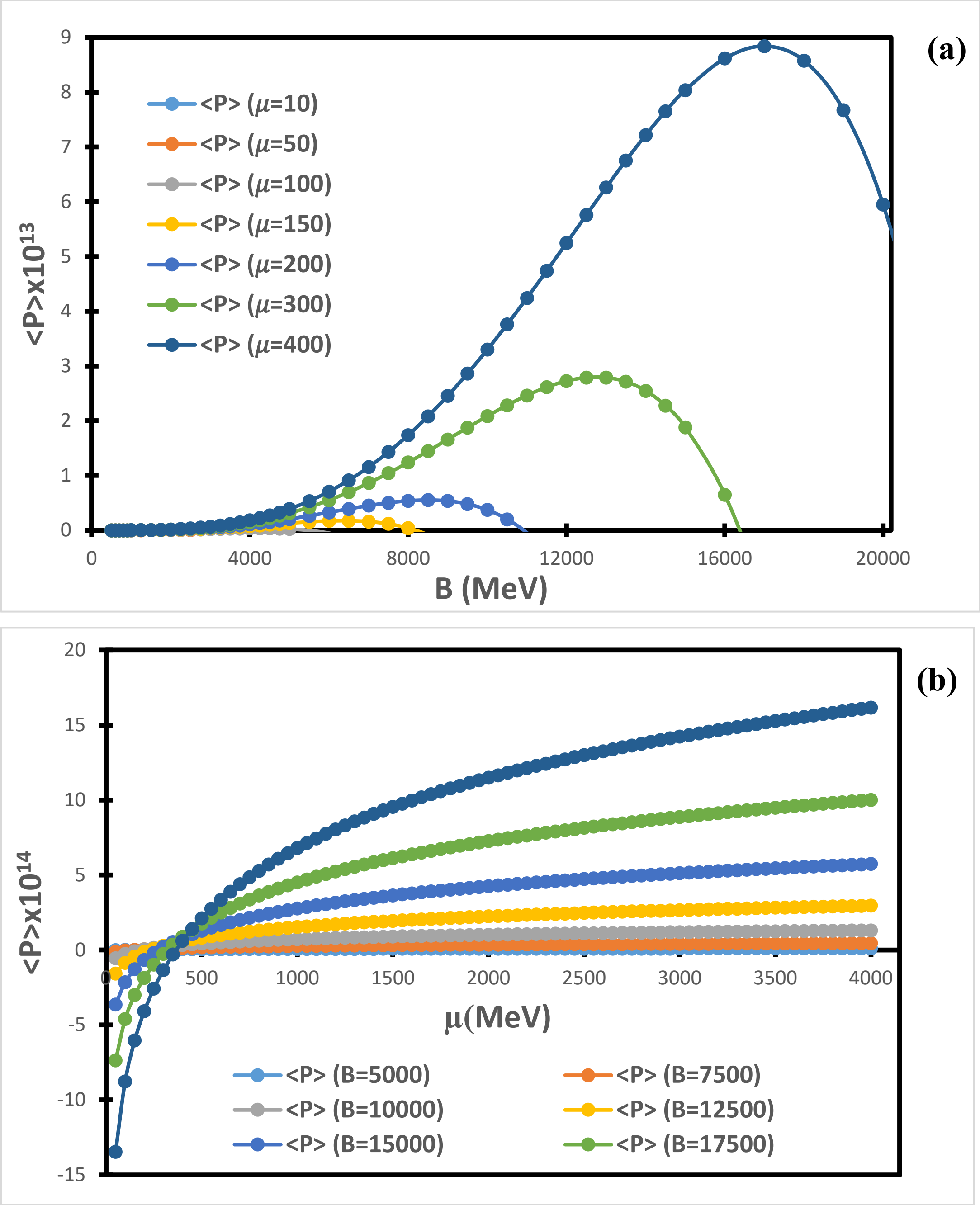

**Fig.1. (a)** Pressure vs magnetic field for different $\mu$ values and **(b)** Pressure vs $\mu$ for different $B$ values

The dependence of EOS of fermion on the statistical parameters shows that the EOS of a system can be used to determine the phase of matter as well because electrons are considered to be in thermal equilibrium with the entire matter in the system. Alternatively, for an ignorable chemical potential ($\mu \ll T$) of the kinetic theory of gases, $\mu$ is just replaced by $T$. These results are comparable to the classical results and justify to use of kinetic gas model for the early universe and the condensed matter model for the interior of NSs. It is worth mentioning that the results, Eqs. (8) and (9), can also be evaluated for $\mu < T$ using the corresponding integrals $I_1$ through $I_6$ from the Appendix of Ref.[13]. However, in that case, the effect of magnetic field and density may not be relevant everywhere. For the very early universe, high $T$ calculations were already done to reveal the properties of electrons [18-20]. Electromagnetic properties of a NS are highly modified due to the properties of statistical medium and can further be investigated through its relation of incoming radiation with the medium. On the other hand, due to existence of active nuclear matter, nuclear interaction is to be incorporated but is postponed in this paper due to its complexity. However, that study is not possible analytically and numerical methods are to be used. However, even the appropriate tools for such type of detailed study are still being developed.

## Appendix:

$$I_1 = \int_M^\infty E^2\, n_F(E,\mu) dE = \frac{M^2}{\beta} a(M\beta,\mu) + \frac{2M}{\beta^2} c(M\beta,\mu) + \frac{2}{\beta^3} d(M\beta,\mu)$$

$$I_2 = \int_M^\infty n_F(E,\mu) dE = \frac{1}{\beta} a(M\beta,\mu)$$

$$I_3 = \int_M^\infty \frac{n_F(E,\mu)}{E^2} dE = \beta g(M\beta,\mu)$$

$$I_4 = \int_M^\infty \frac{n_F(E,\mu)}{E} dE = b(M\beta,\mu)$$

$$I_5 = \int_M^\infty E^3\, n_F(E,\mu) dE = \frac{M^3}{\beta} a(M\beta,\mu) + \frac{3M^2}{\beta^2} c(M\beta,\mu) + \frac{M}{\beta^3} d(M\beta,\mu) + \frac{1}{\beta^4} f(M\beta,\mu)$$

$$I_6 = \int_M^\infty E n_F(E,\mu) dE = \frac{M}{\beta} a(M\beta,\mu) + \frac{1}{\beta^2} c(M\beta,\mu)$$

In the absence of magnetic field, for $\mu > T$ and $\mu \gg m$, the functions are defined as [12];

$$a(m\beta,\mu) = \mu - m - \sum_{l=0}^{\infty}\frac{(-1)^l}{\mu^l}\sum_{n=1}^{\infty}\frac{(-1)^n}{(n\beta)^{1-l}}e^{-n\beta(m-\mu)} = \mu - m$$

$$b(m\beta,\mu) = \sum_{l=0}^{\infty}\frac{(-1)^l}{\mu^l}\sum_{n=1}^{\infty}(-1)^n e^{-n\beta\mu}Ei(-nm\beta) = \ln(\mu/m)$$

$$c(m\beta,\mu) = \frac{\mu^2 - m^2}{2} - \sum_{l=0}^{\infty}\frac{(-1)^l}{\mu^l}\sum_{n=1}^{\infty}\frac{(-1)^n}{(n\beta)^{l}}e^{-n\beta(m-\mu)} = \frac{\mu^2 - m^2}{2}$$

$$d(m\beta,\mu) = \frac{\mu^3 - m^3}{3} - \sum_{l=0}^{\infty}\frac{(-1)^l}{\mu^l}\sum_{n=1}^{\infty}\frac{(-1)^n}{(n\beta)^{l+1}}e^{-n\beta(m-\mu)} = \frac{\mu^3 - m^3}{3}$$

$$f(m\beta,\mu) = \frac{\mu^4 - m^4}{4} - \sum_{l=0}^{\infty}\frac{(-1)^l}{\mu^l}\sum_{n=1}^{\infty}\frac{(-1)^n}{(n\beta)^{l+2}}e^{-n\beta(m-\mu)} = \frac{\mu^4 - m^4}{4}$$

$$g(m\beta,\mu) = \frac{\mu^{-3} - m^{-3}}{3} - \sum_{l=0}^{\infty}\frac{(-1)^l}{\mu^l}\sum_{n=1}^{\infty}\frac{(-1)^n}{(n\beta)^{l+3}}e^{-n\beta(m-\mu)} = \frac{\mu^{-3} - m^{-3}}{3}$$

$$h(m\beta,\mu) = \frac{\mu^{-4} - m^{-4}}{4} - \sum_{l=0}^{\infty}\frac{(-1)^l}{\mu^l}\sum_{n=1}^{\infty}\frac{(-1)^n}{(n\beta)^{l+4}}e^{-n\beta(m-\mu)} = \frac{\mu^{-4} - m^{-4}}{4}$$

The first terms in the functions above vanish for antiparticles (as they yield a negative Fermi energy) and the remaining terms are the same except for $\mu \to -\mu$. However free antiparticles cannot exist in such conditions in a physical system. Using those functions, in the absence of $B$, the number density of electrons can be expressed as;

$$N = \frac{1}{2\pi^2}\left[\frac{\mu^3 - m^3}{3} - \frac{m^2}{2}(\mu - m) - \frac{m^4}{8}\left(\frac{1}{\mu} - \frac{1}{m}\right)\right]$$

The corresponding equation for the energy density is;

$$N < E >= \frac{1}{2\pi^2}\left[\frac{m^3}{2\beta}(\mu - m) - \frac{m^4}{8}\ln\left(\frac{\mu}{m}\right) + \frac{m^2}{2\beta^2}\left(\frac{\mu^2 - m^2}{2}\right) + \frac{m}{\beta}\left(\frac{\mu^3 - m^3}{3}\right) + \frac{1}{\beta^4}\left(\frac{\mu^4 - m^4}{4}\right)\right]$$